\RequirePackage{lineno}
\documentclass[aps,prd,twocolumn,showpacs,superscriptaddress,showpacs]{revtex4}  
\usepackage{epsfig}
\usepackage{graphicx} 
\usepackage{dcolumn}  
\usepackage{bm}       
\usepackage{amssymb}  

\def\mevc {\ifmmode {\rm MeV}/c \else MeV$/c$\fi}
\def\mevcc {\ifmmode {\rm MeV}/c^2 \else MeV$/c^2$\fi}
\def\gevc {\ifmmode {\rm GeV}/c \else GeV$/c$\fi}
\def\gevcc {\ifmmode {\rm GeV}/c^2 \else GeV$/c^2$\fi}
\def\tevcc {\ifmmode {\rm TeV}/c^2 \else TeV$/c^2$\fi}

\def\vtd  {\ifmmode |V_{td}| \else $|V_{td}|$\fi}
\def\vtb  {\ifmmode |V_{tb}| \else $|V_{tb}|$\fi}
\def\vts  {\ifmmode |V_{ts}| \else $|V_{ts}|$\fi}
\def\vcb  {\ifmmode |V_{cb}| \else $|V_{cb}|$\fi}

\newcommand{\xs} {\ifmmode x_{\mbox{\sl s}}
                       \else $x_{\mbox{\sl s}}$\fi}
\newcommand{\xd} {\ifmmode x_d \else $x_d$\fi}
\newcommand{\lxy} {\ifmmode L_{\rm xy} \else $L_{\rm xy}$\fi}
\newcommand{\dgam} {\ifmmode \Delta\Gamma \else $\Delta\Gamma$\fi}
\newcommand{\dm} {\ifmmode \Delta m \else $\Delta m$\fi}
\newcommand{\ctau} {\ifmmode c\tau \else $c\tau$\fi}

\def\EtMiss {E_T\!\!\!\!\!\!\!/ \;\; }

\newcommand{\zto}{Z/\gamma^{*}\rightarrow}
\def\xsec {$\sigma_{t\bar{t}}$}

\begin{document}
\bibliographystyle{apsrev}


\title{First Measurement of the Ratio $\sigma_{t\bar{t}} / \sigma_{Z/\gamma^{*}\rightarrow ll}$ and Precise Extraction of the $t\bar{t}$ Cross Section \\} 
\affiliation{Institute of Physics, Academia Sinica, Taipei, Taiwan 11529, Republic of China} 
\affiliation{Argonne National Laboratory, Argonne, Illinois 60439} 
\affiliation{University of Athens, 157 71 Athens, Greece} 
\affiliation{Institut de Fisica d'Altes Energies, Universitat Autonoma de Barcelona, E-08193, Bellaterra (Barcelona), Spain} 
\affiliation{Baylor University, Waco, Texas  76798} 
\affiliation{Istituto Nazionale di Fisica Nucleare Bologna, $^{cc}$University of Bologna, I-40127 Bologna, Italy} 
\affiliation{Brandeis University, Waltham, Massachusetts 02254} 
\affiliation{University of California, Davis, Davis, California  95616} 
\affiliation{University of California, Los Angeles, Los Angeles, California  90024} 
\affiliation{University of California, San Diego, La Jolla, California  92093} 
\affiliation{University of California, Santa Barbara, Santa Barbara, California 93106} 
\affiliation{Instituto de Fisica de Cantabria, CSIC-University of Cantabria, 39005 Santander, Spain} 
\affiliation{Carnegie Mellon University, Pittsburgh, PA  15213} 
\affiliation{Enrico Fermi Institute, University of Chicago, Chicago, Illinois 60637}
\affiliation{Comenius University, 842 48 Bratislava, Slovakia; Institute of Experimental Physics, 040 01 Kosice, Slovakia} 
\affiliation{Joint Institute for Nuclear Research, RU-141980 Dubna, Russia} 
\affiliation{Duke University, Durham, North Carolina  27708} 
\affiliation{Fermi National Accelerator Laboratory, Batavia, Illinois 60510} 
\affiliation{University of Florida, Gainesville, Florida  32611} 
\affiliation{Laboratori Nazionali di Frascati, Istituto Nazionale di Fisica Nucleare, I-00044 Frascati, Italy} 
\affiliation{University of Geneva, CH-1211 Geneva 4, Switzerland} 
\affiliation{Glasgow University, Glasgow G12 8QQ, United Kingdom} 
\affiliation{Harvard University, Cambridge, Massachusetts 02138} 
\affiliation{Division of High Energy Physics, Department of Physics, University of Helsinki and Helsinki Institute of Physics, FIN-00014, Helsinki, Finland} 
\affiliation{University of Illinois, Urbana, Illinois 61801} 
\affiliation{The Johns Hopkins University, Baltimore, Maryland 21218} 
\affiliation{Institut f\"{u}r Experimentelle Kernphysik, Karlsruhe Institute of Technology, D-76131 Karlsruhe, Germany} 
\affiliation{Center for High Energy Physics: Kyungpook National University, Daegu 702-701, Korea; Seoul National University, Seoul 151-742, Korea; Sungkyunkwan University, Suwon 440-746, Korea; Korea Institute of Science and Technology Information, Daejeon 305-806, Korea; Chonnam National University, Gwangju 500-757, Korea; Chonbuk National University, Jeonju 561-756, Korea} 
\affiliation{Ernest Orlando Lawrence Berkeley National Laboratory, Berkeley, California 94720} 
\affiliation{University of Liverpool, Liverpool L69 7ZE, United Kingdom} 
\affiliation{University College London, London WC1E 6BT, United Kingdom} 
\affiliation{Centro de Investigaciones Energeticas Medioambientales y Tecnologicas, E-28040 Madrid, Spain} 
\affiliation{Massachusetts Institute of Technology, Cambridge, Massachusetts  02139} 
\affiliation{Institute of Particle Physics: McGill University, Montr\'{e}al, Qu\'{e}bec, Canada H3A~2T8; Simon Fraser University, Burnaby, British Columbia, Canada V5A~1S6; University of Toronto, Toronto, Ontario, Canada M5S~1A7; and TRIUMF, Vancouver, British Columbia, Canada V6T~2A3} 
\affiliation{University of Michigan, Ann Arbor, Michigan 48109} 
\affiliation{Michigan State University, East Lansing, Michigan  48824}
\affiliation{Institution for Theoretical and Experimental Physics, ITEP, Moscow 117259, Russia} 
\affiliation{University of New Mexico, Albuquerque, New Mexico 87131} 
\affiliation{Northwestern University, Evanston, Illinois  60208} 
\affiliation{The Ohio State University, Columbus, Ohio  43210} 
\affiliation{Okayama University, Okayama 700-8530, Japan} 
\affiliation{Osaka City University, Osaka 588, Japan} 
\affiliation{University of Oxford, Oxford OX1 3RH, United Kingdom} 
\affiliation{Istituto Nazionale di Fisica Nucleare, Sezione di Padova-Trento, $^{dd}$University of Padova, I-35131 Padova, Italy} 
\affiliation{LPNHE, Universite Pierre et Marie Curie/IN2P3-CNRS, UMR7585, Paris, F-75252 France} 
\affiliation{University of Pennsylvania, Philadelphia, Pennsylvania 19104}
\affiliation{Istituto Nazionale di Fisica Nucleare Pisa, $^{ee}$University of Pisa, $^{ff}$University of Siena and $^{gg}$Scuola Normale Superiore, I-56127 Pisa, Italy} 
\affiliation{University of Pittsburgh, Pittsburgh, Pennsylvania 15260} 
\affiliation{Purdue University, West Lafayette, Indiana 47907} 
\affiliation{University of Rochester, Rochester, New York 14627} 
\affiliation{The Rockefeller University, New York, New York 10021} 
\affiliation{Istituto Nazionale di Fisica Nucleare, Sezione di Roma 1, $^{hh}$Sapienza Universit\`{a} di Roma, I-00185 Roma, Italy} 

\affiliation{Rutgers University, Piscataway, New Jersey 08855} 
\affiliation{Texas A\&M University, College Station, Texas 77843} 
\affiliation{Istituto Nazionale di Fisica Nucleare Trieste/Udine, I-34100 Trieste, $^{ii}$University of Trieste/Udine, I-33100 Udine, Italy} 
\affiliation{University of Tsukuba, Tsukuba, Ibaraki 305, Japan} 
\affiliation{Tufts University, Medford, Massachusetts 02155} 
\affiliation{Waseda University, Tokyo 169, Japan} 
\affiliation{Wayne State University, Detroit, Michigan  48201} 
\affiliation{University of Wisconsin, Madison, Wisconsin 53706} 
\affiliation{Yale University, New Haven, Connecticut 06520} 
\author{T.~Aaltonen}
\affiliation{Division of High Energy Physics, Department of Physics, University of Helsinki and Helsinki Institute of Physics, FIN-00014, Helsinki, Finland}
\author{J.~Adelman}
\affiliation{Enrico Fermi Institute, University of Chicago, Chicago, Illinois 60637}
\author{B.~\'{A}lvarez~Gonz\'{a}lez$^v$}
\affiliation{Instituto de Fisica de Cantabria, CSIC-University of Cantabria, 39005 Santander, Spain}
\author{S.~Amerio$^{dd}$}
\affiliation{Istituto Nazionale di Fisica Nucleare, Sezione di Padova-Trento, $^{dd}$University of Padova, I-35131 Padova, Italy} 

\author{D.~Amidei}
\affiliation{University of Michigan, Ann Arbor, Michigan 48109}
\author{A.~Anastassov}
\affiliation{Northwestern University, Evanston, Illinois  60208}
\author{A.~Annovi}
\affiliation{Laboratori Nazionali di Frascati, Istituto Nazionale di Fisica Nucleare, I-00044 Frascati, Italy}
\author{J.~Antos}
\affiliation{Comenius University, 842 48 Bratislava, Slovakia; Institute of Experimental Physics, 040 01 Kosice, Slovakia}
\author{G.~Apollinari}
\affiliation{Fermi National Accelerator Laboratory, Batavia, Illinois 60510}
\author{A.~Apresyan}
\affiliation{Purdue University, West Lafayette, Indiana 47907}
% ...

\collaboration{CDF Collaboration\footnote{With visitors from $^a$University of Massachusetts Amherst, Amherst, Massachusetts 01003,
$^b$Universiteit Antwerpen, B-2610 Antwerp, Belgium, 
$^c$University of Bristol, Bristol BS8 1TL, United Kingdom,
$^d$Chinese Academy of Sciences, Beijing 100864, China, 
$^e$Istituto Nazionale di Fisica Nucleare, Sezione di Cagliari, 09042 Monserrato (Cagliari), Italy,
$^f$University of California Irvine, Irvine, CA  92697, 
$^g$University of California Santa Cruz, Santa Cruz, CA  95064, 
$^h$Cornell University, Ithaca, NY  14853, 
$^i$University of Cyprus, Nicosia CY-1678, Cyprus, 
$^j$University College Dublin, Dublin 4, Ireland,
$^k$University of Edinburgh, Edinburgh EH9 3JZ, United Kingdom, 
$^l$University of Fukui, Fukui City, Fukui Prefecture, Japan 910-0017
$^m$Kinki University, Higashi-Osaka City, Japan 577-8502
$^n$Universidad Iberoamericana, Mexico D.F., Mexico,
$^o$University of Iowa, Iowa City, IA  52242,
$^p$Kansas State University, Manhattan, KS 66506
$^q$Queen Mary, University of London, London, E1 4NS, England,
$^r$University of Manchester, Manchester M13 9PL, England,
$^s$Muons, Inc., Batavia, IL 60510, 
$^t$Nagasaki Institute of Applied Science, Nagasaki, Japan, 
$^u$University of Notre Dame, Notre Dame, IN 46556,
$^v$University de Oviedo, E-33007 Oviedo, Spain, 
$^w$Texas Tech University, Lubbock, TX  79609, 
$^x$IFIC(CSIC-Universitat de Valencia), 56071 Valencia, Spain,
$^y$Universidad Tecnica Federico Santa Maria, 110v Valparaiso, Chile,
$^z$University of Virginia, Charlottesville, VA  22906
$^{aa}$Bergische Universit\"at Wuppertal, 42097 Wuppertal, Germany,
$^{bb}$Yarmouk University, Irbid 211-63, Jordan
$^{jj}$On leave from J.~Stefan Institute, Ljubljana, Slovenia, 
}}
\noaffiliation


\date{\today}

\begin{abstract}

We report a measurement of the ratio of the $t\bar{t}$ to $Z/\gamma^{*}$ production cross sections in $\sqrt{s}=1.96$ TeV $p\bar{p}$ collisions using data corresponding to an integrated luminosity of up to 4.6 $\mathrm{fb}^{-1}$, collected by the CDF II detector.  The $t\bar{t}$ cross section ratio is measured using two complementary methods, a $b$-jet tagging measurement and a topological approach.  By multiplying the ratios by the well-known theoretical $Z/\gamma^{*}\rightarrow ll$ cross section predicted by the standard model, the extracted $t\bar{t}$ cross sections are effectively insensitive to the uncertainty on luminosity.  A best linear unbiased estimate is used to combine both measurements with the result  $\sigma_{t\bar{t}} = 7.70 \pm 0.52 \; \; \mathrm{pb}$, for a top-quark mass of 172.5 GeV/$c^2$.



\end{abstract}


\maketitle

We describe two measurements of the $t\bar{t}$ cross section (\xsec), one based on $b$-jet tagging, where backgrounds are reduced using a $b$-hadron identification technique, and the other a topological approach, which uses event kinematics to distinguish $t\bar{t}$ events from backgrounds.  Measurements of the $t\bar{t}$ cross section test perturbative QCD at high energy, and serve as a probe for possible new physics \cite{uwer}.  Because of the top quark's unusually large mass compared to other fermions, it is possible that the top quark plays some special role in electroweak symmetry breaking \cite{r1}.  This new physics can manifest as an enhancement, or even deficit, in the rate of top quark pair production.  Measurements of the $t\bar{t}$ cross section serve as tests of these possible new physics processes and can place stringent limits on these models.

Previous related cross section measurements have uncertainties larger than ten percent and have used less than or equal to an integrated luminosity of 1 $\mathrm{fb}^{-1}$ \cite{tagxsec,annxs,doxsec}.  The measurements presented in this paper use up to 4.6 $\mathrm{fb}^{-1}$ of collected data, enough to be limited by systematic uncertainties.  The largest systematic uncertainty for both measurements results from the uncertainty on the integrated luminosity.  To reduce the luminosity uncertainty on the $t\bar{t}$ cross section measurement, the $Z/\gamma^{*}\rightarrow ll$ cross section is measured in the same corresponding data sample and the ratio of the $t\bar{t}$ to $Z/\gamma^{*}\rightarrow ll$ cross sections calculated.  The $t\bar{t}$ cross section is determined by multiplying the ratio by the theoretical $Z/\gamma^{*}\rightarrow ll$ cross section predicted by the standard model.  This replaces a 6$\%$ uncertainty from the measured luminosity with a 2$\%$ uncertainty from the theoretical $Z/\gamma^{*}\rightarrow ll$ cross section.  This is the first application of this technique to a $t\bar{t}$ cross section measurement, and the combination of the two $t\bar{t}$ cross section measurements has a precision of 7$\%$.


Events are collected at the Collider Detector Facility (CDF) at Fermi National Accelerator Laboratory \cite{CDF,CDF2}.  The components relevant to these cross section measurements include the silicon tracker, the central outer tracker (COT), the electromagnetic and hadronic calorimeters, the muon detectors, and the luminosity counters.


At the Tevatron, the top quark is expected to be produced mostly in pairs through quark antiquark annihilation and gluon fusion \cite{uwer}.  Assuming unitarity of the three-generation CKM matrix, top quarks decay almost exclusively to a $W$-boson and a bottom quark.  Because of this, the signature of $t\bar{t}$ events in the detector is determined by how the $W$ bosons decay.  The analyses presented here identify $t\bar{t}$ events using the decay of one $W$-boson to quarks and the other to a lepton and a neutrino.  

Candidate $t\bar{t}$ events are first collected through central high-$p_{T}$ lepton triggers \cite{CDF2,coordinates}.  Each event is required to have a single high-$p_{T}$ electron or muon.  Tau-lepton reconstruction has lower purity and therefore taus are not specifically selected, though some events pass selection when a tau decays leptonically.  Electrons are required to be central and have a track in the COT along with a large clustered energy deposit in the electromagnetic calorimeter ($E_{T} > 20$ GeV and $|\eta| < 1.1$), with little energy in the hadronic calorimeter.  Muons are required to have a high-$p_{T}$ track in the COT ($p_{T} > 20$ GeV and $|\eta| < 0.6$), a small amount of minimum-ionizing energy in the calorimeters, and associated set of hits in the muon detectors.  Events are required to have a large amount of missing transverse energy as evidence of a neutrino from the $W$-boson decay: $\EtMiss > 25$(35) GeV for the $b$-jet tagging (topological) measurement \cite{metdef}.  At least three reconstructed jets are required, where a jet is identified using a fixed cone algorithm of radius $R = \sqrt{(\Delta \eta)^2 + (\Delta \phi)^2} = 0.4 $ \cite{jes}.  Each jet is required to have transverse energy $E_T > 20$ GeV and $|\eta| < 2$.  To reduce contamination by background processes, the $b$-jet tagging measurement requires at least one identified $b$-quark jet, in which some tracks in the jet are found to come from a secondary vertex, displaced from the primary vertex, due to the longer lifetime of a $b$-hadron \cite {g4topxsec}.  To further reduce background, an additional requirement is placed on the scalar sum ($H_{T}$) of the transverse energy of the lepton, $\EtMiss$, and jets ($H_{T} > 230$ GeV) for the $b$-jet tagging measurement.  


There are several physics processes which can mimic a $t\bar{t}$ event in the selected data sample, such as $W$+jets, $Z$+jets, diboson ($WW$, $ZZ$, $WZ$), electroweak produced top-quarks (single-top), and QCD multijet processes.  The $b$-jet tagging and topological measurements differ in their approaches to reducing and normalizing these backgrounds.  We first discuss the $b$-jet tagging and the topological measurements, and then the $Z/\gamma^{*}\rightarrow ll$ cross section and ratio.

The $b$-jet tagging measurement uses a mixture of data and Monte Carlo (MC) techniques to estimate the contribution of each process.  Backgrounds are initially calculated before requiring a $b$-tagged jet (pretag), and the predicted number of $b$-tagged events is then derived from the pretag estimate.  For the pretag prediction, $Z$+jets, diboson, and single top quark events are generated using {\sc alpgen}, {\sc pythia}, and {\sc madevent} respectively, where {\sc pythia} is used to model parton showering and the underlying event for all generated samples \cite{alpgen1,alpgen2,alpgen3,pythia,madgraph}.   CTEQ6.6 parton distribution functions (PDF) are used in all MC simulations \cite{cteq66}.  {\sc cdfsim}, a {\sc geant}-based simulation, is used to model the CDF detector response \cite{cdfsim,geant}.  The $Z$+jets, diboson, and single top quark samples are normalized to their respective theoretical cross sections \cite{diboson,singletop}.  QCD multijet background is difficult to model using MC simulations, and therefore a data-driven approach is taken, which is described in the cited literature \cite{antielectron}.  Acceptance of $t\bar{t}$ events is modeled by {\sc pythia}, where the pole-mass of the top quark ($M_{t}$) is set to $172.5$ GeV/$c^2$.  The $t\bar{t}$ cross section, for the pretag estimate, is preliminarily set to the standard model expectation \cite{uwer}.  The contribution from $W$+jets is normalized to the total number of pretag events in data minus the estimate for $t\bar{t}$, QCD multijet, diboson, single-top, and $Z$+jets events.  



With the pretag estimate for all processes in hand, the number of events with at least one $b$-tagged jet for $Z$+jets, diboson, and single-top events is found by applying a MC-based tagging efficiency to all pretag estimates.  For $W$+jets, the relative fraction of jets associated with heavy flavor (HF) is found to be under-predicted in the MC simulation.  A correction factor, to be applied on all the W+jets samples, is obtained using the experimental data by measuring the W + HF content in W plus single jet events, using an artificial neural network (ANN) trained to discriminate HF from light flavor (LF) jets, and comparing it to the prediction for the corresponding simulated samples \cite{ann,kit}.  The number of $W$ plus HF events with at least one $b$-tagged jet is estimated by applying this correction factor and a tagging efficiency to the predicted number of pretag $W$+HF events.  Events with a $W$ boson associated with LF jets enter into the data sample when a jet is wrongly identified as a HF jet (mistagged jet).  This is the result of poorly reconstructed tracks in the detector which happen to form a displaced secondary vertex, and is difficult to model in the simulation.  Instead, the probability that a jet is mistagged is determined using independent multi-jet data and parameterized by $E_{T}$, $\eta$, $\phi$, number of tracks in the jet, and sum of the $E_{T}$ in the detector.  The fraction of mistagged events in the $b$-tagged data sample is found by applying the mistag parameterization to the pretag data.  The number of QCD multijet events with a $b$-tagged jet are calculated in the same manner as the pretag multijet estimate.




\begin{figure}[t]
    \begin{center}
     \includegraphics[scale=0.42]{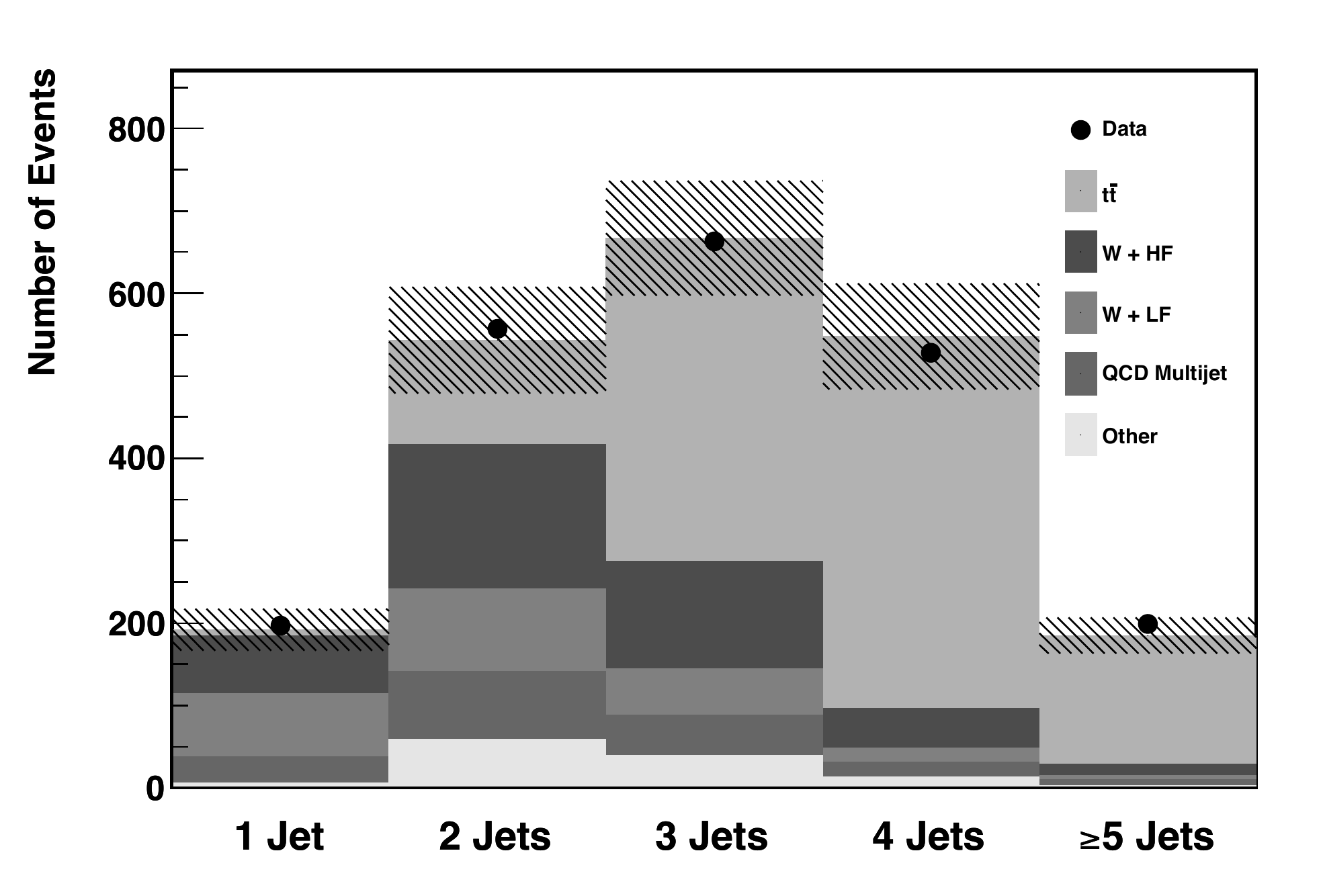}
      \caption{Number of data and predicted background events as a function of jet multiplicity, with the number of $t\bar{t}$ events normalized to the measured cross section.  The hashed lines represent the uncertainty on the predicted number of events.  \label{njetplot}}
    \end{center}
\end{figure}


To measure the $t\bar{t}$ cross section, a likelihood is formed from the data, the $t\bar{t}$ cross section, and the predicted background for that cross section.  Using collected data corresponding to an integrated luminosity of 4.3 $\mathrm{fb}^{-1}$, the result is $\sigma_{t\bar{t}} = 7.22 \pm 0.35_{stat} \pm 0.56_{sys} \pm 0.44_{lum} \; \; \mathrm{pb}$.  The predicted number of events for each background process, along with the number of expected $t\bar{t}$ events at the measured cross section, is shown compared to data in Figure \ref{njetplot}.  The largest systematic uncertainties, shown in Table \ref{systematics}, come from the measured luminosity, the correction to the $W$ + HF background, and the $b$-tag modeling in the simulation. 





\begin{figure}[t]
    \begin{center}
     \includegraphics[scale=0.40]{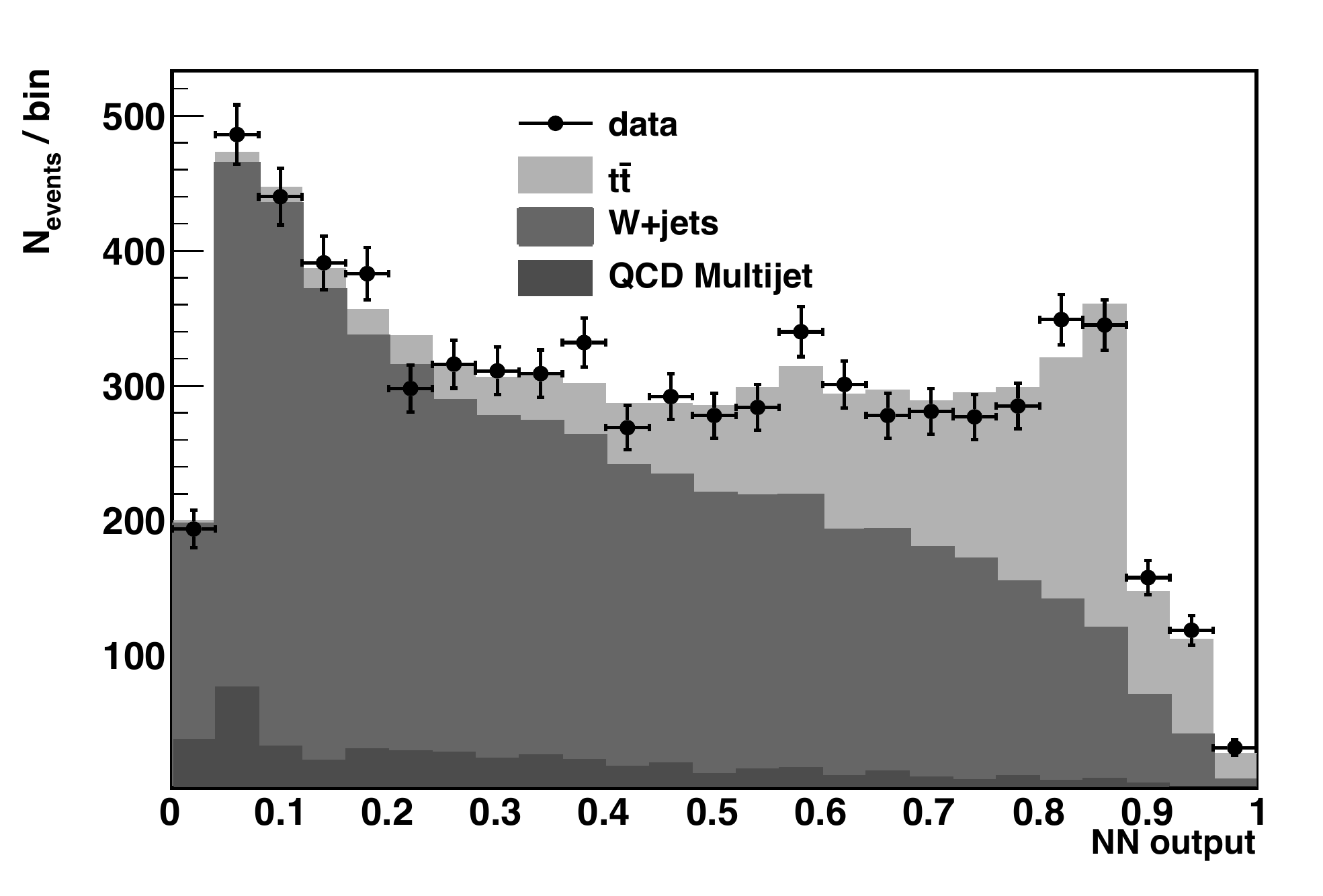}
      \caption{
      The output of an artificial neural network (ANN), trained to distinguish $t\bar{t}$ events from background, for simulated $t\bar{t}$ and background events, and data.  The $t\bar{t}$ cross section is extracted from a fit of templates to the data. \label{annoutput}}
    \end{center}
\end{figure}

The topological measurement uses an ANN to discriminate $t\bar{t}$ events from background by exploiting differences in their kinematics \cite{ann}.  Because of the large mass of the top quark, $t\bar{t}$ events are more energetic, central, and isotropic compared with the dominant backgrounds such as $W$+jets and QCD multijet events, whose kinematics are more influenced by the boost from the momentum distribution of the colliding partons.  To exploit these kinematic differences, seven different kinematic distributions are used as an input to an ANN:  $H_T$; the aplanarity \cite{aplanarity} of the event; $\sum p_{Z} / \sum E_{T}$ of jets; $\sum E_{T}$ of jets excluding the two highest $E_{T}$; minimum invariant mass between 4-vectors of any two jets; minimum angle between any two jets; and the maximum $|\eta|$ of any jet.  $W$+jets events are the dominant background process in the pretag data sample, and therefore the ANN is trained using only $t\bar{t}$ and $W$+jets simulated samples.  Templates of the ANN output distributions are obtained from {\sc pythia} $t\bar{t}$ and {\sc alpgen} $W$+jets MC samples, as well as the same data-derived model for QCD multijet background as in the $b$-jet tagging measurement.  The templates are fit to the ANN output distribution of data events.  The absolute normalizations of the $W$+jets and $t\bar{t}$ distributions are considered unknown and allowed to float in the fit.  The QCD multijet normalization is obtained using a similar method to the $b$-jet tagging measurement.  The templates are used in a binned likelihood fit of the ANN output to extract the $t\bar{t}$ cross section.  Figure \ref{annoutput} shows the output of the ANN for signal and background templates fit to the data.
 
Using collected data corresponding to an integrated luminosity of 4.6 $\mathrm{fb}^{-1}$, the result of the topological measurement is $\sigma_{t\bar{t}} = 7.71 \pm 0.37_{stat} \pm 0.36_{sys} \pm 0.45_{lum} \; \; \mathrm{pb}$.  Because $b$-tagging is not used in this measurement, it is insensitive to two of the largest sources of systematic uncertainty of the $b$-jet tagging measurement, as shown in Table \ref{systematics}.



The large luminosity uncertainty on the $t\bar{t}$ cross section measurements, which is due to the uncertainty on the inelastic $p\bar{p}$ cross section and acceptance of the luminosity counters, can be effectively removed by measuring them relative to the inclusive $Z/\gamma^{*}\rightarrow ll$ cross section, and multiplying by the theoretical $Z/\gamma^{*}\rightarrow ll$ cross section.  The uncertainties on the theoretical and measured $Z/\gamma^{*}\rightarrow ll$ cross sections are propagated to the final $t\bar{t}$ cross section measurement, but are small compared to the luminosity uncertainty.

The inclusive $Z/\gamma^{*}\rightarrow ll$ cross section is measured using consistent trigger requirements and lepton identification with the corresponding $t\bar{t}$ cross section measurement so that the integrated luminosity is the same.  Because silicon tracking is not always active during detector operation, the $b$-jet tagging measurement uses a slightly smaller integrated luminosity than the topological measurement.  Therefore, the $Z/\gamma^{*}\rightarrow ll$ cross section is measured for two non-identical data samples.

Events are selected using two oppositely charged electrons or muons with an invariant mass ($M_{l\bar{l}}$) between 66 and 116 GeV/$c^2$.  The $Z/\gamma^{*}\rightarrow ll$ signal acceptance is modeled by an inclusive {\sc pythia} MC where $Z/\gamma^{*}$ decays to $e^{{\rm -}} \quad e^{+}$ and \textit{$\mu$}$^{{\rm -}}$ \textit{$\mu $}$^{+}$ final states.  Although the $Z/\gamma^{*}\rightarrow ll$ process is a clean signal, there are some small backgrounds from diboson, $t\bar{t}$, $W$ + jet, and $Z/\gamma^{*}\rightarrow ll$ events from outside the mass range.  Diboson and $t\bar{t}$ contributions are modeled from inclusive {\sc pythia} MC and fixed to their respective theoretical cross sections \cite{diboson,uwer}.   A small number of QCD multijet and $W$+jets events pass through selection when at least one jet is misreconstructed as a lepton.  We estimate this contribution by studying like-charge events in data that pass our event selection.


The measured cross section times branching ratio for $Z/\gamma^{*}\rightarrow ll$ events in the invariant mass range of 66-116 GeV/$c^2$ is $\sigma_{\zto\ell\ell} = 247.8 \pm 0.8_{stat} \pm 4.4_{sys} \pm 14.6_{lum} \;\; \mathrm{pb}$ for the integrated luminosity used in both the $b$-jet-tagging and topological measurements.  This is consistent with the standard model prediction $\sigma_{Z/\gamma^{*}\rightarrow ll} = 251.3 \pm 5.0 \; \mathrm{pb}$ \cite{CDF2}.  The largest systematic uncertainty on the measured $Z/\gamma^{*}\rightarrow ll$ cross section comes from the measured luminosity, as shown in Table \ref{systematics}.


\begin{table}[t]
\footnotesize
\caption{Systematic uncertainties ($\Delta \sigma / \sigma$ $\%$) on the measured $t\bar{t}$ and $Z/\gamma^{*}\rightarrow ll$ cross sections.  Several uncertainties are reduced in the ratio ($\sigma_{t\bar{t}} / \sigma_{Z/\gamma^{*}\rightarrow ll}$) due to correlations between the measurements.  \label{systematics}}
\begin{center} 
\begin{tabular}{cccc}
\hline
\hline
Systematic & $t\bar{t}_{tag}$ &  $t\bar{t}_{ANN}$ & $Z/\gamma^{*}\rightarrow ll$\\
\hline
Luminosity & 6.1  & 5.8 & 5.9  \\
$b$-tag modeling & 4.7 & - & - \\
$W$+HF correction & 4.0 & - & -\\
Jet energy scale & 4.1 & 2.9 & -\\
Monte Carlo generator & 2.7 & 2.6 & -\\
Initial/final state radiation & 0.6 & 0.4 & - \\
PDF &  0.6 & 0.9 & 1.4\\
Background shape model & 0.2 & 1.9 & 0.3 \\
Lepton ID/trigger & 1.3 & 1.3  & 1.1 \\

\hline
Total & 10.0 & 7.5 & 6.2  \\
Total  $\sigma_{t\bar{t}} / \sigma_{Z/\gamma^{*}\rightarrow ll}$ & 8.2  & 4.7  & \\
\hline
\hline
\end{tabular}
\end{center}
\end{table}


The measured ratio of the $t\bar{t}$ to $Z/\gamma^{*}\rightarrow ll$ cross sections for the $b$-tagging (topological) measurement is $2.77 \pm 0.15_{stat} \pm 0.25_{sys} \; \%$ ($3.12 \pm 0.15_{stat} \pm 0.16_{sys} \; \%$).  Multiplying this ratio by the theoretical $Z/\gamma^{*}\rightarrow ll$ cross section, the $t\bar{t}$  cross sections using $b$-tagging and event topologies are $\sigma_{t\bar{t}} = 7.32 \pm 0.36_{stat} \pm 0.59_{sys} \pm 0.14_{theory} \; \; \mathrm{pb}$ and $\sigma_{t\bar{t}} = 7.82 \pm 0.38_{stat} \pm 0.37_{sys} \pm 0.15_{theory} \; \; \mathrm{pb}$, respectively.  The luminosity systematic uncertainty for both measurements has been replaced by a small uncertainty from the theoretical $Z/\gamma^{*}\rightarrow ll$ cross section.  The correlations between the uncertainties in lepton identification, trigger efficiencies, and parton distribution functions  for the  $t\bar{t}$ and $Z/\gamma^{*}\rightarrow ll$ cross section measurements are positive and have been taken into account in the ratio.  As jets are not used in the measurement of the $Z/\gamma^{*}\rightarrow ll$ cross section, all other systematic uncertainties are found to be independent.  


The two measurements are combined using a best linear unbiased estimate \cite{blue1}.  A covariance matrix is constructed from statistical and systematic uncertainties for each result.  The matrix is inverted to extract a weight for each of the two results, and the results are combined using the corresponding weight.  The combined cross section for $t\bar{t}$  production is $\sigma_{t\bar{t}} = 7.70 \pm 0.52 \; \; \mathrm{pb}$ for a top quark mass $M_{t} = 172.5$ GeV/$c^2$.  The result is consistent with the standard model next-to-leading order prediction  $\sigma_{t\bar{t}} = 7.45^{+0.72}_{-0.63} \; \; \mathrm{pb}$ \cite{uwer}.


\begin{acknowledgments}
We thank the Fermilab staff and the technical staffs of the participating institutions for their vital contributions. This work was supported by the U.S. Department of Energy and National Science Foundation; the Italian Istituto Nazionale di Fisica Nucleare; the Ministry of Education, Culture, Sports, Science and Technology of Japan; the Natural Sciences and Engineering Research Council of Canada; the National Science Council of the Republic of China; the Swiss National Science Foundation; the A.P. Sloan Foundation; the Bundesministerium f\"ur Bildung und Forschung, Germany; the Korean Science and Engineering Foundation and the Korean Research Foundation; the Science and Technology Facilities Council and the Royal Society, UK; the Institut National de Physique Nucleaire et Physique des Particules/CNRS; the Russian Foundation for Basic Research; the Comisi\'on Interministerial de Ciencia y Tecnolog\'{\i}a, Spain; the European Community's Human Potential Programme; the Slovak R\&D Agency; and the Academy of Finland.
\end{acknowledgments}

\bibliography{ratio}

\begin{thebibliography}{26}
\bibitem[{coo()}]{coordinates}
\bibinfo{note}{CDF uses a cylindrical coordinate system with the $z$ axis along
  the proton beam axis. Pseudorapidity is $\eta \equiv -\ln(\tan (\theta/2))$,
  where $\theta$ is the polar angle, and $\phi$ is the azimuthal angle relative
  to the proton beam direction, while $p_T = |p|\ \sin(\theta)$, $E_T = E\
  \sin(\theta)$.}
\bibitem[{met()}]{metdef}
\bibinfo{note}{Missing transverse energy, $\EtMiss$, is defined as the
  magnitude of the vector $ -\sum_i E_T^i \vec n_i$ where $E_T^i$ are the
  magnitudes of transverse energy contained in each calorimeter tower $i$, and
  $\vec n_i$ is the unit vector from the interaction vertex to the tower in the
  transverse ($x,y$) plane.}
\end{thebibliography}



\end{document}